\documentclass[english,conference, 10pt]{IEEEtran}
\usepackage[T1]{fontenc}
\usepackage[latin9]{inputenc}
\usepackage{float}
\usepackage{amsthm}
\usepackage{amsmath}
\usepackage{amssymb}
\usepackage{graphicx}
\usepackage{esint}

\makeatletter

\floatstyle{ruled}
\newfloat{algorithm}{tbp}{loa}
\providecommand{\algorithmname}{Algorithm}
\floatname{algorithm}{\protect\algorithmname}

\makeatother

\usepackage{babel}
\begin{document}

\title{Training Optimization for Energy Harvesting Communication Systems}

\author{Yaming Luo, Jun Zhang, and Khaled B. Letaief, \textit{Fellow, IEEE}\\
ECE Department, Hong Kong University of Science and Technology, Hong
Kong\\
Email: \{luoymhk, eejzhang, eekhaled\}@ust.hk}
\maketitle
\begin{abstract}
Energy harvesting (EH) has recently emerged as an effective way to
solve the lifetime challenge of wireless sensor networks, as it can
continuously harvest energy from the environment. Unfortunately, it
is challenging to guarantee a satisfactory short-term performance
in EH communication systems because the harvested energy is sporadic.
In this paper, we consider the channel training optimization problem
in EH communication systems, i.e., how to obtain accurate channel
state information to improve the communication performance. In contrast
to conventional communication systems, the optimization of the training
power and training period in EH communication systems is a coupled
problem, which makes such optimization very challenging. We shall
formulate the optimal training design problem for EH communication
systems, and propose two solutions that adaptively adjust the training
period and power based on either the instantaneous energy profile
or the average energy harvesting rate. Numerical and simulation results
will show that training optimization is important in EH communication
systems. In particular, it will be shown that for short block lengths,
training optimization is critical. In contrast, for long block lengths,
the optimal training period is not too sensitive to the value of the
block length nor to the energy profile. Therefore, a properly selected
fixed training period value can be used. 
\end{abstract}

\section{Introduction}

In traditional wireless sensor networks, the limited energy at each
node constrains the network  lifetime. Energy harvesting (EH) is a
promising technology which has the potential to provide a powerful
solution to achieve perpetual lifetime without requiring external
power cables or periodic battery replacement \cite{key-1}. Energy
harvesting nodes can harvest energy from the environment, including
solar energy, vibration energy, thermoelectric energy, RF energy,
etc. With its highly self-reliance capability, EH will undoubtedly
 play an important role in future green communication networks.

However, employing energy harvesting nodes poses new challenges related
to the link and network design, as the harvested energy is typically
small and random. Thus although EH technology improves the long-term
performance, the challenging \textit{short-term performance} needs
to be guaranteed. Previous works on EH networks have developed communication
protocols to either maximize the throughput or minimize the transmission
completion time, assuming perfect channel state information (CSI)
at the transmitter and receiver, e.g., \cite{key-2,key-3,key-4}.
In \cite{key-3}, a directional water-filling (DWF) algorithm is proposed
to solve the transmit power allocation problem in EH systems, while
in \cite{key-4}, a generalized DWF algorithm is proposed to solve
a general utility maximization problem.

In a wireless communication link, CSI is important, e.g., for the
receiver to decode the transmitted message, or for rate adaptation
at the transmitter. At the receiver side, CSI can be obtained by sending
pilot symbols from the transmitter. There exists a tradeoff between
the training overhead and the training performance. Specifically,
spending too much energy or time on channel training will reduce the
energy or time for data transmission. On the other hand, training
with  too little energy or time will degrade the estimation performance.
To maximize the throughput, the training period and training power
should be carefully selected. Previous studies have shown that in
conventional communication systems, training power optimization and
training period optimization are \textit{decoupled}, of which the
\textit{power optimization} is more important. In \cite{key-5}, it
was shown that for a point-to-point link without the peak power constraint,
the optimal training policy involves sending one pilot symbol with
optimized training power. However, in EH communication systems, the
training design is different and is largely influenced by the low
rate and randomness property of the available energy. The selection
of the training period and training power in EH systems are coupled
and both will depend on the EH profile in the communication block.
Therefore, the training design in EH communication systems is more
challenging and plays a more important role.

In this paper, we investigate the training optimization problem in
EH communication systems. We first characterize the properties of
the training design in an EH communication system. We then propose
two different training policies to determine the training period and
power. The first training policy adaptively adjusts the training period
based on the energy profile in the whole transmission block, while
the second one is designed in an adaptive way according to the average
EH rate of the block. Simulation results will show that training optimization
is important to improve the communication performance in EH systems,
especially when the transmission block is not very long. For long
block lengths, the optimal training period is not too sensitive to
the value of the block length. Therefore, a fixed training period
value can be used if properly selected.

\section{System Model}

We consider a point-to-point communication link where the transmitter
is an EH node, as shown in Figure 1. The transmitter can only use
the energy it harvests, and we assume that all the harvested energy
is used for communication. The channel is characterized by block fading,
and within a coherence block, the channel gain $h$ is constant with
$h\sim\textrm{CN}\left(0,\sigma_{h}^{2}\right)$. The additive white
Gaussian noise is denoted as $n$ with $n\sim\textrm{CN}\left(0,\sigma^{2}\right)$.
The communication within one transmission block includes two stages:
the training stage and the data transmission stage. The partition
of the two stages is in the unit of a time slot $T_{S}$. The fading
block length is denoted as $T$, with $N=\frac{T}{T_{S}}$ slots,
while the training stage length is $T_{t}$, with $N_{t}=\frac{T_{t}}{T_{S}}$
slots. During the training stage, the receiver obtains an estimate
of $h$, denoted as $\hat{h}$, through the use of a pilot signal.
The estimation error is denoted as $\tilde{h}$ with $\tilde{h}=h-\hat{h}$.
Before the transmission stage, the receiver feeds back the value of
$\hat{h}$ to the transmitter. The feedback channel is assumed to
be perfect, while the case with unideal feedback will be discussed
in future work.

\begin{figure}
\includegraphics[scale=1.3]{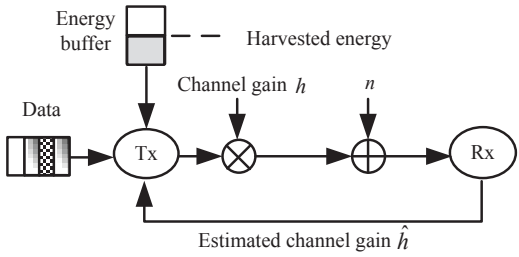}

\caption{The basic system model.}
\end{figure}

\subsection{Energy Model}

An important factor that determines the performance of an EH system
is the \textit{EH profile}, which models the variation of the harvested
energy with time. Several different types of EH profiles are shown
in the left part of Figure 2. For convenience, we plot all EH profiles
inside a 2-D coordinate system of accumulated energy versus time.

To demonstrate the property and impact of energy profiles, we adopt
similar EH assumptions as in \cite{key-2,key-3}. Specifically, we
assume that the energy profile in the considered transmission block
is known before the communication starts. This assumption is applicable
for predictable energy models, such as solar energy \cite{key-6}.

The utilization of the harvested energy is constrained by the EH profile,
and therefore the \textit{energy neutrality constraint} exists in
EH systems \cite{key-7}. The energy neutrality means that the energy
consumed thus far cannot exceed the total energy harvested. For simplicity,
we assume that the EH node can only use the energy harvested in the
previous slots. If the consumed power is denoted as $P\left(t\right)$,
the initial energy in the energy buffer as $E_{0}$, and the harvested
energy in the $k$th slot as $E_{k}$, then the energy neutrality
constraints can be expressed as 
\begin{equation}
\int_{0}^{l{T_{S}}}{P}\left(t\right)dt\le\sum_{k=0}^{l-1}E_{k},
\end{equation}
where $l$ is the index of the time slot with $l=1,2,...,N$. 

As shown in (1), a certain EH profile determines a \textit{feasible
energy consumption domain}, and only the policies inside this domain
are\textit{ feasible energy consumption policies}, both of which are
plotted in the same coordinate system with the EH profile in the right
part of Figure 2. Due to the energy neutrality constraint (1), we
cannot use the energy arriving in the future, but can back up the
current energy for future use. This causal energy constraint determines
the directional property of all power allocation policies in EH systems,
which will be discussed in more detail later.

Among all kinds of EH processes, there are two special cases: the
\textit{non-EH} case and the \textit{constant-rate EH} case, as shown
in the left part of Figure 2. Here we treat the conventional non-EH
system, i.e., without the EH function and only with the average power
constraint, as an extreme case of energy harvesting, in which all
the energy arrives before the first slot. This is equivalent to relaxing
all the causal energy constraints. The feasible energy domain of non-EH
nodes is the union of all the possible EH profiles with the same total
energy in a given time duration, so it provides the \textit{best}
performance among all the EH profiles. Constant-rate EH refers to
the node that can harvest energy at a constant rate. In this case,
the profile can be considered as a deterministic process. In practical
systems, when the EH profile does not change frequently or the block
length is small, a constant-rate EH profile is a good approximation
of the energy profile in each transmission block, with the mean of
the EH process as its harvesting rate.

\begin{figure}
\includegraphics[scale=0.35]{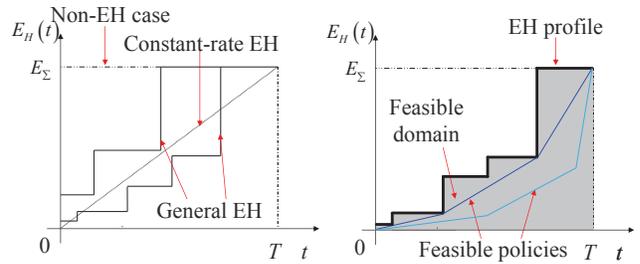}

\caption{The energy profile and feasible energy consumption policies in the
2-D coordinate system of accumulated energy versus time. The left
part plots energy profiles, for two general EH cases, and two special
cases: the non-EH case and the constant-rate EH case. The right part
plots the feasible energy consumption domain and policies for a given
EH profile, of which the bold line is the EH profile, the area under
it is the feasible domain, and a curve connecting the bottom-left
point and the top-right point inside this domain is a feasible energy
consumption policy, two examples of which are shown.}
\end{figure}

The battery capacity is also an important factor for the EH link performance
besides the EH profile. In this paper we assume that the energy buffer
is of an infinite capacity, while the case with a finite buffer capacity
will be dealt with in future work.

\section{Impact of Channel Training in EH systems}

In this section, we first investigate the training policy for EH systems
and compare it with non-EH systems. We will then develop power allocation
for the data transmission stage.

\subsection{Training Stage in EH Systems}

In the training stage, we denote the average training power in the
$j$th time slot as $\bar{P_{j}}$ $\left(1\leq j\leq N_{t}\right)$,
then the variance of the estimation error with an MMSE channel estimator
is \cite{key-8}\textbf{\vspace{-0.1in}}

\begin{equation}
\sigma_{\tilde{h}}^{2}=\frac{\sigma^{2}\sigma_{h}^{2}}{\sigma^{2}+\sigma_{h}^{2}\sum_{j=1}^{N_{t}}\bar{P_{j}}}.
\end{equation}

We see that only the sum of average training powers matters. This
means that as long as the total training power is the same, the training
performance is fully determined, independent of the training period
or the power allocation inside this stage. Thus, we will use the discrete-time
expression of $P_{j}=\bar{P_{j}}$ to denote training powers. 

Due to the causal energy constraint in EH systems, there exists a
big difference in the training design for non-EH systems and EH systems.
In the non-EH system without a peak power constraint, the optimal
$N_{t}$ is always 1 \cite{key-5}, as shown in the left part of Figure
3. An intuitive explanation is that we can always achieve a good training
performance with enough training power (as long as it is less than
the total power available). Meanwhile, we shall make the training
period as small as one time slot. Thus, what matters is the power
allocated for channel training rather than the training period. However,
this is not the case for the EH system. Due to the stochastic EH profile,
the energy arrival in the first time slot may be very small, as shown
in the right part of Figure 3. Hence, fixing the training period as
1 slot will generally provide an inaccurate channel estimation. The
total training power is largely determined by the training period,
which makes it more important than the power allocation, and increases
the difficulty of the training design. 

In EH systems, we select such a training power allocation policy that,
for a given $N_{t}$, all the harvested energy for $1\leq j\leq N_{t}-1$
is exhausted, while there may be some energy left at slot $N_{t}$,
of which the value is optimized. This is \textit{optimal} because
it is not possible to find a smaller training period $N_{t}'<N_{t}$
to achieve the same training performance.

\begin{figure}
\includegraphics[scale=0.35]{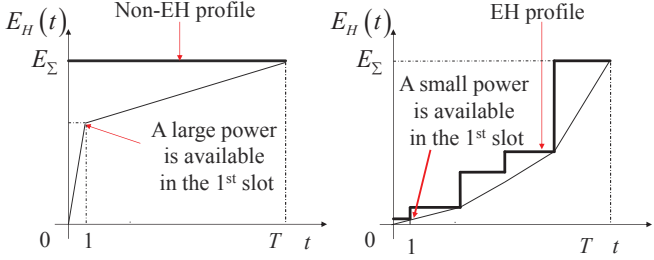}

\caption{Comparison of the power allocation policies in the training stage
for non-EH and EH systems. The left and right figures represent  the
non-EH and EH cases, respectively. }
\end{figure}

\subsection{Data Transmission Stage in EH Systems with Estimation Errors}

Considering the channel estimation error and the training overhead,
the average achievable throughput in each time slot is \textbf{\vspace{-0.2in}}

\begin{equation}
R=\textrm{E}_{\hat{h}}\left[\frac{1}{N}\sum_{i=N_{t}+1}^{N}\log_{2}\left(1+\frac{\left|\hat{h}\right|^{2}P_{i}}{\sigma^{2}+P_{i}\sigma_{\tilde{h}}^{2}}\right)\right].
\end{equation}
As shown in \cite{key-9}, this is a lower bound for the capacity
with channel estimation error and we will use it as the performance
metric in the paper.

By substituting (3) and adopting (9) in \cite{key-10}, this rate
expression can be finally transformed to \textbf{\vspace{-0.0in}}

\begin{equation}
R=\frac{1}{N}\log_{2}\left(e\right)\sum_{i=N_{t}+1}^{N}\exp\left(\frac{1}{K_{i}}\right)\textrm{E}_{1}\left(\frac{1}{K_{i}}\right),
\end{equation}
where $K_{i}=\frac{\sigma_{h}^{4}P_{i}\sum_{j=1}^{N_{t}}P_{j}}{\sigma^{4}+\sigma^{2}\sigma_{h}^{2}P_{i}+\sigma^{2}\sigma_{h}^{2}\sum_{j=1}^{N_{t}}P_{j}}$,
and $\textrm{E}_{1}\left(x\right)=\int_{x}^{+\infty}t^{-1}e^{-t}dt$.

Different from non-EH systems that use a constant transmit power in
the data transmission stage, in the EH system, we need to determine
the power allocation between different time slots, as the power allocated
to each slot needs to satisfy the energy neutrality constraint (1).
For given $N_{t}$, $\hat{h}$ and $\sigma_{\tilde{h}}^{2}$, the
power allocation problem is as follows:\medskip{}

\noindent \textbf{Problem 1:}

\noindent 
\[
\begin{array}{cl}
\max\limits _{P_{N_{t}+1},...P_{N}} & \frac{1}{N}\sum_{i=N_{t}+1}^{N}\log_{2}\left(1+\frac{\left|\hat{h}\right|^{2}P_{i}}{\sigma^{2}+P_{i}\sigma_{\tilde{h}}^{2}}\right)\\
{\rm {s}}.{\rm {t}}. & {T_{S}}\sum\limits _{i=N_{t}+1}^{l}{P_{i}}\le E_{te}+\sum_{k=N_{t}}^{l-1}E_{k}\\
 & {T_{S}}\sum\limits _{i=N_{t}+1}^{N}{P_{i}}=E_{te}+\sum_{k=N_{t}}^{N-1}E_{k}\\
 & l=N_{t}+1,...,N-1,
\end{array}
\]
where $E_{te}$ denotes the energy left from the training operation,
and is known before the optimization.

In Problem 1, the first constraint is the energy neutrality constraint.
In contrast to non-EH systems, even if the channel stays unchanged,
the power still needs to be adaptively allocated from slot to slot
due to the causal EH constraints. The second constraint means that
at the end of the block, the node needs to use up all the available
energy, as we do not consider the energy sharing between blocks to
render our problem tractable, while the case with block-to-block energy
sharing will be discussed in future work.

We make the following two comments on Problem 1. First, similar to
the training power, the data transmission power is expressed in a
discrete-time form, as it is optimal for the power inside one slot
to be constant due to the concavity of the objective function. Second,
as the training power and the data transmission power are the same
from the perspective of energy consumption, we use the same notation
$P$ and only distinguish between them by the time index. 

The throughput expression with estimation errors satisfies the condition
of the directional water-filling (DWF) algorithm \cite{key-4}, and
thus the optimal power allocation follows DWF. Such a DWF algorithm
has a special property that the solution is only determined by constraints,
irrespective of the parameters in the objective function. So in our
problem the solution is independent of $\hat{h}$ and $\sigma_{\tilde{h}}^{2}$,
i.e., the estimation performance and the value of the estimated channel
gain do not have any impact on the power allocation. This special
property can largely simplify the power allocation, as there is no
need to completely reallocate the data transmission power for different
values of $N_{t}$. We only need to execute the power allocation over
the whole block once for $N_{t}=0$, and update a few points for other
values of $N_{t}$. Accordingly, we develop an efficient algorithm
to solve Problem 1 for different $N_{t}$, as shown in Algorithm 1.

\begin{algorithm}
\begin{enumerate}
\item Initialization: Set integers $k_{0}=0$ and $n=1$.
\item Iteration: Iterate $k_{n}=\arg\underset{k:k\leq N}{\min}\left\{ \frac{\sum_{j=k_{n-1}}^{k-1}E_{j}}{k-k_{n-1}}\right\} $
with $n$ adding 1 each time, until $k{}_{n}=N$, so finally an index
set $\mathcal{K}_{0}=\left\{ k_{n}\right\} $ is constructed.
\item Results for $N_{t}$=0: The optimal power in the $i$th slot is $p_{n}=\frac{\sum_{j=k_{n-1}}^{k_{n}-1}E_{j}}{k_{n}-k_{n-1}}$
for $i\in\left[k_{n-1}+1,k_{n}\right]$, and a power set $\mathcal{P}_{0}=\left\{ p_{n}\right\} $
is obtained for $N_{t}$=0. 
\item Update for $N_{t}\neq0$: Reset $k'_{0}=N_{t}$, recalculate $k'_{1}=\arg\underset{k:k\leq N}{\min}\left\{ \frac{\sum_{j=k'_{0}}^{k-1}E_{j}}{k-k'_{0}}\right\} $,
then the index set for $N_{t}$ is $\mathcal{K}_{N_{t}}=\left\{ k'_{1}\right\} \cup\left\{ \textrm{all }k_{n}\in\mathcal{K}_{0}\textrm{ that }k_{n}>k'_{1}\right\} $.
\item Results for $N_{t}\neq0$: The power for $i\in\left[k_{0}+1,k'_{1}\right]$
is $p'_{1}=\frac{\sum_{j=k'_{0}}^{k'_{1}-1}E_{j}}{k'_{1}-k'_{0}}$,
while the other powers are unchanged, then the power set for $N_{t}$
is $\mathcal{P}_{N_{t}}=\left\{ p'_{1}\right\} \cup\left\{ \textrm{all }p_{n}\in\mathcal{P}_{0}\textrm{ that }p_{n}>p'_{1}\right\} $.
\end{enumerate}
\caption{DWF algorithm for Problem 1 with different $N_{t}$}
\end{algorithm}

From Algorithm 1, we can see that for a given $N_{t}$, the power
allocation result consists of several intervals, and the power is
a constant value inside each of these intervals. The endpoint indices
of all intervals form an\textit{ index set }$\mathcal{K}_{N_{t}}$,
while the powers in these intervals form a \textit{power allocation
set }$\mathcal{P}_{N_{t}}$. Furthermore, according to Steps 4 and
5 of Algorithm 1, for different $N_{t}$, the majority ($i\in\left[k'_{1},N\right]$)
of the transmit power allocation is unvaried, while only a \textit{small}
proportion ($i\in\left[k_{0}+1,k'_{1}\right]$) changes with $N_{t}$.
This property brings the possibility of decoupling the training power
allocation and the selection of training period, which will be used
in the next section for the optimal training design.

\section{Optimal Training Design}

As seen from the last section, in EH communication systems, the coupling
of the training period selection and the training power allocation
brings the main difficulty in the training design, and the training
period selection is especially critical. In this section, we will
investigate the optimal training design in EH systems and propose
two training policies.

\subsection{Problem Formulation}

With the average throughput in (4) as the objective and considering
energy neutrality constraints, the optimal training problem in EH
communication systems is formulated as \smallskip{}
\smallskip{}
\smallskip{}

\noindent \textbf{Problem 2:\vspace{-0.3in}}

\[
\begin{array}{cl}
\max\limits _{P_{1},...P_{N},N_{t}} & \sum\limits _{i=N_{t}+1}^{N}{\exp}\left({\frac{1}{{K_{i}}}}\right){{\rm {E}}_{1}}\left({\frac{1}{{K_{i}}}}\right)\\
{\rm {s}}.{\rm {t}}. & 1\le{N_{t}}<N,\begin{array}{c}
\end{array}N_{t}\in\mathbb{N}\\
 & {T_{S}}\sum\limits _{i=1}^{l}{P_{i}}\le\sum_{k=0}^{l-1}E_{k},\begin{array}{c}
\end{array}l=1,2,...,N-1\\
 & {T_{S}}\sum\limits _{i=1}^{N}{P_{i}}=\sum_{k=0}^{N-1}E_{k}.
\end{array}
\]
\textbf{\vspace{-0.1in}}

This problem has two difficulties: 1) the optimization of $N_{t}$
and $P_{i}$ are coupled; 2) the optimization variable $N_{t}$ exists
in the summation limit in the objective function, and only takes discrete
values. Due to the intractability of this problem, we propose a sub-optimal
solution in the next subsection.

\subsection{A Sub-optimal Solution}

Due to the difficulties of Problem 2, we adopt a DWF approximation
and a rate approximation to derive a sub-optimal solution. Both of
these simplifications have good approximation properties, which will
be verified by the simulation results. In addition, for the special
case of the constant-rate EH, both approximations become equivalent
to the original problem. As commented in Section II, the constant-rate
EH model is a good approximation for the energy profile in each transmission
block in different EH systems, so our sub-optimal solution will be
in general close to optimal.

\subsubsection{DWF Approximation}

First, based on the property of Algorithm 1 as discussed in Section
III.B, we make an approximation to decouple the training power allocation
and the training period selection. From Algorithm 1, the power allocation
in the whole transmission block only changes in a small number of
slots for different values of $N_{t}$. We make an approximation that
the power allocation is fixed for all values of $N_{t}$, i.e., we
ignore the possible changes of power allocation in some slots for
different $N_{t}$. This simplification will decouple $N_{t}$ and
$P_{i}$, so that we can perform the DWF power allocation just once,
and then optimize $N_{t}$ over a fixed power allocation result. In
this way, we can get a sub-optimal solution with low computational
complexity.

\subsubsection{Rate Approximation}

With the DWF approximation, the problem is still intractable, as the
variable $N_{t}$ only takes integer values and appears in the summation
of the objective function. To further simplify the problem, we make
the following rate approximation: first, for a given value  of $N_{t}$,
we calculate the estimation error assuming a constant training power
to equal the average EH rate $\bar{P}_{H}$, i.e., $\sigma_{\tilde{h}}^{2}=\frac{\sigma^{2}\sigma_{h}^{2}}{\sigma^{2}+\sigma_{h}^{2}\sum_{j=1}^{N_{t}}P_{j}}\approxeq\frac{\sigma^{2}\sigma_{h}^{2}}{\sigma^{2}+\sigma_{h}^{2}N_{t}\bar{P}_{H}}$;
second, we determine the achievable throughput $\hat{R}$ assuming
all the slots including the training period are used for data transmission
with the transmit power equal to the DWF result in the respective
slot, i.e., $\hat{R}=\sum_{i=1}^{N}M_{i}$, where $M_{i}=\exp\left(\frac{1}{K_{i}}\right)\textrm{E}_{1}\left(\frac{1}{K_{i}}\right)$
is the average throughput for the $i$th slot considering the estimation
error; finally, we include the throughput loss due to the training
period, i.e., $R=\sum_{i=N_{t}+1}^{N}M_{i}\approxeq\frac{N-N_{t}}{N}\sum_{i=1}^{N}M_{i}=\frac{N-N_{t}}{N}\hat{R}$.
To summarize, Step 1 is to consider the effect of the estimation error,
Step 2 is adopting the DWF approximation while ignoring the time taken
by training, and Step 3 is to take the time consumed by training back
into consideration. While greatly simplifying the problem, this rate
approximation preserves the essential tradeoff in the original training
design problem, i.e., the tradeoff between the resource consumed by
training and the estimation performance.

\subsubsection{Solution to the Simplified Problem }

Based on previous two steps, the training design problem can be formulated
as: 

\noindent \textbf{\vspace{-0.15in}}

\noindent \textbf{Problem 3:}

\noindent \textbf{\vspace{-0.33in}}

\[
\begin{array}{ll}
\underset{_{N_{T}}}{\max} & \frac{N-N_{t}}{N}\sum_{i=1}^{N}\exp\left(\frac{1}{K_{Si}}\right)\textrm{E}_{1}\left(\frac{1}{K_{Si}}\right)\\
{\rm {s}}.{\rm {t}}. & 1\le{N_{t}}<N,\begin{array}{c}
\end{array}N_{t}\in\mathbb{N},
\end{array}
\]

\textbf{\vspace{-0.09in}}

\noindent where $K_{Si}=\frac{\sigma_{h}^{4}P_{i}N_{t}\bar{P}_{H}}{\sigma^{4}+\sigma^{2}\sigma_{h}^{2}P_{i}+\sigma^{2}\sigma_{h}^{2}N_{t}\bar{P}_{H}}$,
and all $P_{i}$ are determined through Step 1\textasciitilde{}3 of
Algorithm 1. The solution for Problem 3 is a sub-optimal solution
for Problem 2.

For simplicity, we denote $x=\frac{1}{N_{t}}$. When $x$ is assumed
continuous, the objective function is concave, the proof of which
is omitted due to space limitation. Through the derivative with respect
to $x$ we can get an implicit solution, i.e., the solution of Problem
3 is the solution of the following equation (the discretization part
is omitted due to space limitation)\textbf{\vspace{-0.15in}}

\begin{equation}
\sum_{i=1}^{N}\left\{ M_{i}\left[1+\left(\frac{N}{N_{t}}-1\right)\frac{\sigma^{2}}{\sigma_{h}^{2}P_{i}G_{i}}\right]-\frac{\left(\frac{N}{N_{t}}-1\right)}{1+G_{i}}\right\} =0,
\end{equation}
\textbf{\vspace{-0.2in}}

\noindent where $G_{i}=\frac{\sigma_{h}^{2}\bar{P}_{H}N_{t}}{\sigma^{2}+\sigma_{h}^{2}P_{i}}$.

When $N$ approaches infinity, we can get an asymptotic solution of
$N_{t}$ and its ratio over $N$ in closed form as \textbf{\vspace{-0.124in}}

\begin{equation}
N_{t}=\frac{2N}{1+\sqrt{1+4NW}},\begin{array}{c}
\end{array}\,\begin{array}{c}
\end{array}\,\begin{array}{c}
\end{array}\,\begin{array}{c}
\end{array}\,\alpha=\frac{N_{t}}{N}=\frac{2}{1+\sqrt{1+4NW}},
\end{equation}
\textbf{\vspace{-0.05in}}where $W=\frac{\sum_{i=1}^{N}M_{i}^{A}}{\sum_{i=1}^{N}\frac{\sigma^{4}+\sigma^{2}\sigma_{h}^{2}P_{i}}{\sigma^{2}\sigma_{h}^{2}\bar{P}_{H}}\left(1-\frac{\sigma^{2}M_{i}^{A}}{\sigma_{h}^{2}P_{i}}\right)}$,
and $M_{i}^{A}=\exp\left(\frac{\sigma^{2}}{\sigma_{h}^{2}P_{i}}\right)\textrm{E}_{1}\left(\frac{\sigma^{2}}{\sigma_{h}^{2}P_{i}}\right)$.

From the expression of solution (6), we see that the optimal training
period is influenced by the block length and EH profiles. \textit{Generally
speaking, a larger block length will result in a longer training period
$N_{t}$, but a smaller training period ratio $\alpha=\frac{N_{t}}{N}$.
When $N$ approaches infinity, $N_{t}$ also approaches infinity,
while the ratio $\alpha$ approaches zero. }

\subsection{A Special Case -- The Constant-rate EH Profile}

The constant-rate EH process can be used to approximate any general
EH system when the energy harvesting rate does not change intensively.
Thus, in this section we will show that the optimal solution of the
constant-rate EH case can provide another sub-optimal solution for
the general EH systems with the same average EH rate. This solution
is very practical as it only needs the mean value of the EH profile,
rather than its instantaneous realization. 

To optimize $N_{t}$ for the constant-rate EH case, the gradient analysis
of throughput shows that the optimal value for both the training and
transmission powers equals the EH rate, denoted by $P_{H}$, i.e.,
the transmit power is a constant in both stages, and we only need
to determine $N_{t}$, the training period.

By applying (5) to the constant-rate EH case, the optimal $N_{t}$
for the constant-rate EH is the solution of\textbf{\vspace{-0.05in}}
\begin{equation}
M^{Con}\left[1+\left(\frac{N}{N_{t}}-1\right)\frac{\sigma^{2}}{\sigma_{h}^{2}P_{H}G{}^{Con}}\right]-\frac{\left(\frac{N}{N_{t}}-1\right)}{1+G^{Con}}=0,
\end{equation}
where $G^{Con}=\frac{\sigma_{h}^{2}P_{H}N_{t}}{\sigma^{2}+\sigma_{h}^{2}P_{H}}$,
$M^{Con}=\exp\left(\frac{1}{K^{Con}}\right)\textrm{E}_{1}\left(\frac{1}{K^{Con}}\right)$,
and $K^{Con}=\frac{\sigma_{h}^{4}P_{H}^{2}N_{t}}{\sigma^{4}+\sigma^{2}\sigma_{h}^{2}P_{H}\left(1+N_{t}\right)}$.

Note that (7) is the exact optimal solution for a constant-rate EH
system. Meanwhile, it also provides an approximate solution for a
general EH communication system. Thus, we propose a second sub-optimal
solution to Problem 2 as follows. First, we equate the total energy
in a given transmission block of a general EH system with that of
a constant-rate EH system, from which we can get an equivalent $\hat{P}_{H}$,
the average EH rate. Then, the approximate solution is derived by
solving the training optimization problem for a constant-rate EH system
with rate $\hat{P}_{H}$. This provides a sub-optimal value of $N_{t}$.
Once we get this value of $N_{t}$, the directional water-filling
algorithm can be applied for power allocation in the data transmission
stage to further improve the performance.

\section{Simulation Results}

In this section, we provide simulation results to show the importance
of training optimization in EH communication systems. We will compare
the throughput performances of the optimal policy, two sub-optimal
policies, and several fixed training policies. The result of the optimal
policy, i.e., the solution of Problem 2,  is obtained by exhaustive
search. The sub-optimal solutions include (5) as sub-optimal solution
1, and (7) as sub-optimal solution 2. The fixed training policies
include: fixing a training period value $\ensuremath{N_{t}}$, fixing
the training period ratio $\ensuremath{\frac{N_{t}}{N}}$, and the
conventional fixing 1-slot policy, i.e., $N_{t}=1$. These training
policies will also be compared with two performance upper bounds.
One upper bound assumes perfect CSI and with the same EH process.
It will be denoted as ``upper bound 1''. The second is the non-EH
case with the same total energy in each transmission block and adopting
the optimal channel training in \cite{key-5}. We shall denote it
as ``upper bound 2''. 

In the simulation, we assume that the channel is distributed as $h\thicksim\textrm{CN\ensuremath{\left(0,1\right)}}$.
Both the energy arrival process in each time slot and the initial
energy in the energy buffer are assumed to be Poisson distributed,
with parameter $\lambda_{e}$ set to be 1. The average SNR is also
1.The simulation is run for 1000 random EH realizations. We select
$N_{t}=30$ for the fixed training period scheme, and $N_{t}=0.04N$
for the fixed training period ratio scheme. The results are shown
in Figure 4.

We see that the optimal policy and two sub-optimal policies are very
close to each other, and all have small gaps to the performance bounds.
The achievable throughput of sub-optimal solution 2 is slightly lower
than that of sub-optimal solution 1. It should be emphasized that
the sub-optimal solution 2 does not need the instantaneous realization
of the EH process, but only the average energy harvesting rate of
the process, which makes it more practical. We can also find that
when $N$ is small, the gaps between all the fixed policies and the
optimal one are very large, which means that we need to adaptively
adjust the training period for different EH profiles. However, when
$N$ is large, the throughput gaps between the fixed policies and
the optimal one are not very big, except for $N_{t}=1$. This means
that in a low mobility environment, i.e., with a large $N$, it is
feasible to select a fixed training period or ratio not only independent
of the EH process, but also independent of the block length\textit{.}

\begin{figure}
\includegraphics[scale=0.6]{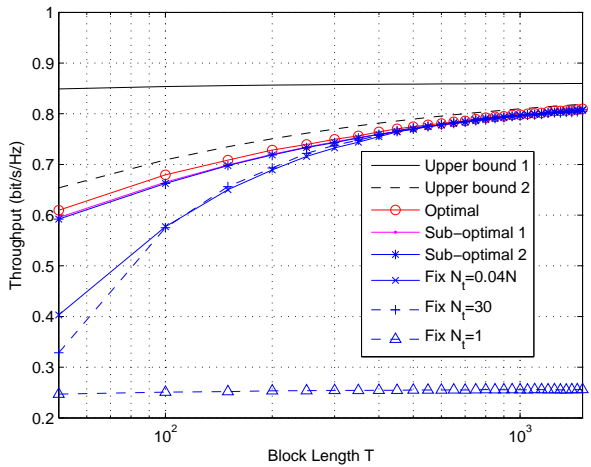}

\caption{The comparison of different training schemes for various block lengths
assuming a Poisson EH process. }
\end{figure}

Next, we elaborate more on the fixed $N_{t}$ policy, as we cannot
change $N_{t}$ in some practical systems. Considering typical values
of coherence bandwidth $W_{c}=500\textrm{kHz}$ and coherence time
$T_{c}=2.5\textrm{ms}$ (from \cite{key-11}) as an example, the block
length is $N=1250$. Figure 5 compares the throughput of the optimal
policy with adaptively selected $N_{t}$ and the fixed policy with
different fixed values of $N_{t}$. We can see that when $N_{t}$
lies in the interval $\left[13,128\right]$, the performance gap between
the fixed policy and optimal policy is within $5\%$; while the interval
for a $10\%$ gap is $\left[8,189\right]$. This indicates that as
long as $N_{t}$ belongs to a proper region, it is a fairly good policy
to fix $N_{t}$ independent of the instantaneous EH process. 

From these results, we see that if the training period can be adaptively
adjusted in each block, we can get the approximate optimal value using
the sub-optimal solution 2 in (7). On the other hand, if the training
period needs to be fixed, we can select a proper fixed value according
to the throughput gap requirement, which will work well especially
for the low mobility environment.\textbf{\vspace{-0.0in}}

\begin{figure}
\includegraphics[scale=0.6]{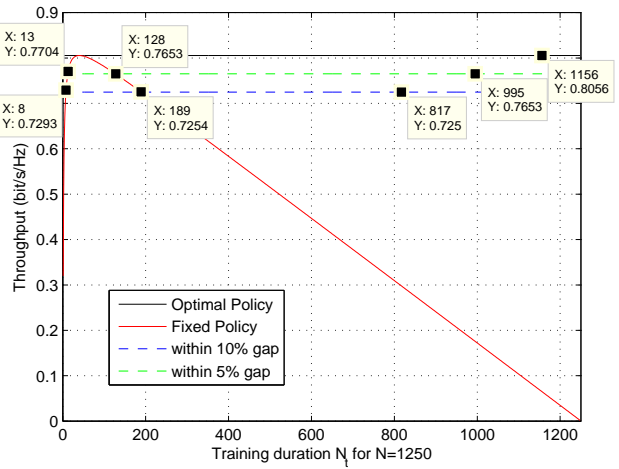}

\caption{Performance of the optimal policy and the fixed policy for $N$ =
1250 with a Poisson EH process. The optimal policy adaptively picks
a value of $N_{t}$ for different energy profiles, while the fixed
policy always chooses a single $N_{t}$.}
\end{figure}

\section{Conclusions\vspace{-0.00in}}

In this paper, we investigated the optimal training design for EH
communication systems, which was shown to be quite different from
conventional non-EH systems and poses new challenges. We found that
the training period should be carefully selected, especially when
the coherence block length is not very long. In particular, we proposed
two sub-optimal training policies to determine the training period
and power, the second of which is especially attractive as it only
requires information about the average EH rate instead of the detailed
energy profile in each transmission block. Furthermore, we demonstrated
that in low mobility environments, a carefully selected fixed training
period can provide satisfactory performance, which provides a practical
option for systems that cannot adaptively adjust the training period.

\end{document}